\documentclass[epj,referee]{svjour}
\usepackage{latexsym}
\usepackage{graphics}
\usepackage{subfig}
\usepackage[percent]{overpic}
\begin{document}
\title{The Accelerating Growth of Online tagging Systems}
\author{Lingfei Wu}
\institute{Dept. of Media and Communication, City University of Hong Kong, Hong Kong, China}
\date{Received: date / Revised version: date}
%
\abstract{Research on the growth of online tagging systems not only is interesting in its own right, but also yields insights for website
management and semantic web analysis. Traditional models that describing the growth of online systems can be divided between linear and nonlinear
 versions. Linear models, including the BA model (Brabasi and Albert, 1999), assume that the average activity of users is a constant independent
  of population. Hence the total activity is a linear function of population. On the contrary, nonlinear models suggest that the average
  activity is affected by the size of the population and the total activity is a nonlinear function of population. In the current study,
  supporting evidences for the nonlinear growth assumption are obtained from data on Internet users' tagging behavior. A power law relationship between the number of new tags ($F$) and the population ($P$), which can be expressed
  as ${F}\sim{P}^{\gamma}$ ($\gamma>1$), is found. I call this pattern accelerating growth and find it relates the to time-invariant
  heterogeneity in individual activities. I also show how a greater heterogeneity leads to a faster growth.} 
\authorrunning {L. Wu}
\titlerunning {The Accelerating Growth of Online Tagging Systems}
\maketitle
\section{Introduction}
\label{intro}

In the research literature, there are two kinds of models describing the growth of online social systems: linear and nonlinear. Linear models suggest that the expected value of individual activities $M$ is a constant independent of active population $P$. As a consequence, the total amount of activity $F = MP$ is a linear function of $P$. For example, in the BA model for social network evolution \cite{1}, the average degree is a constant parameter. On the contrary, nonlinear models argue that the size of the population has an effect on individual activities \cite{2}, which leads to a nonlinear relationship between $F$ and $P$. The assumption of nonlinear growth is supported by findings in different online activities including game playing \cite{3}, resource recommendation \cite{4}, collaborative programming \cite{5} and tagging \cite{6}. Moreover, evidence on the nonlinear nature of human collective behavior is also seen in many offline systems, such as cities and countries \cite{7}\cite{8}, academic groups \cite{9}, sexual networks \cite{10}, and so on.

In the current study, I investigate two online tagging systems and obtain supporting evidence for the nonlinear growth assumption. I find that within a tagging system, the number of new tags $F$ is a power law function of the active population $P$ with a constant exponent $\gamma$ despite the daily fluctuation of tags and population. Since $\gamma>1$ means that the number of new tags grows faster than the population does, the pattern can be named as``accelerating growth" \cite{2}\cite{11}\cite{12}. It is observed that, the growth rate $\gamma$ positively relates to the heterogeneity in individual tagging activities $1/\beta$, in which $\beta$ is the exponent of power law distribution of individual activities (Eq.3). The heterogeneity $1/\beta$ is found to remain constant over time within a system, but differs across systems.

This paper is organized as follows. In section~\ref{sec.2}, I introduce the empirical accelerating growth patterns in two online tagging systems. In section~\ref{sec.3}, I show how the growth rate $\gamma$ relates to the heterogeneity of individual activities $1/\beta$, and validate their relationship by empirical data and simulation. Finally, I briefly summarize the findings and discuss possible applications of accelerating growth in the informational industry.

\section{Accelerating growth of online tagging systems}
\label{sec.2}

\begin{center}
\begin{table*}[ht]\footnotesize
\caption{The empirical values of accelerating growth rate in online tagging systems..}
\hfill{}
\begin{tabular}{lcccccccccc}
\hline
  Activity & System & $\gamma$ &$\theta$&  95$\%$ CI of $\gamma$  & Adjusted $R^{2}$ & N of Days & Web Address\\ \hline
  Book-tagging & Delicious & 1.18 &0.18 &  [1.11, 1.31] & 0.80 &  171 & delicious.com\\
  Photo-tagging & Flickr & 1.39 & 0.39 & [1.32, 1.46] & 0.94 & 120 & flickr.com\\ \hline
\end{tabular}
\hfill{}
\label{tab.1}
\end{table*}
\end{center}

\begin{figure}
\resizebox{1\columnwidth}{!}{
  \includegraphics{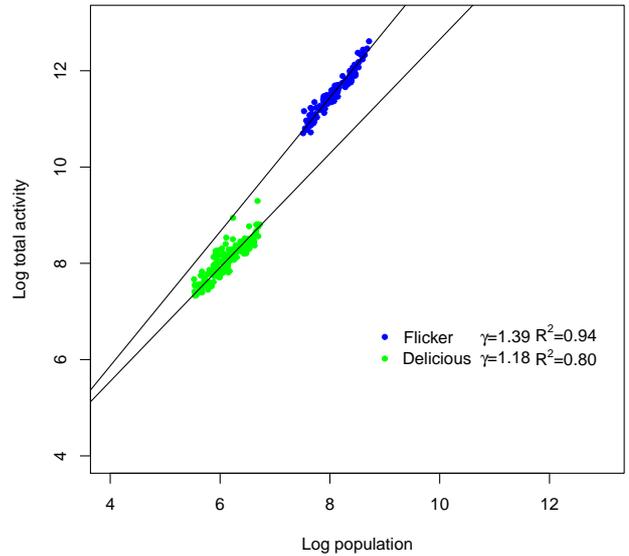}
}
\caption{Accelerating growth in online tagging systems. Different
datasets are marked in points of different colors (blue for
Flickr and green for Delicious).The x axis is the log of active population in a day and the y axis is the log of total activity in the day.
The orthogonal log-log regression lines are also shown.}
\label{fig.1}
\end{figure}

I investigate the tagging activities of 77,917 users on Flickr and 5,147 users on Delicious. The data from Flickr and Delicious are collected by systematically crawling the record of all active users \cite{13}. In the two systems, I define $P$ as the number of active users in a day, and $F$ the number of new tags generated by the users. Please refer to Table~\ref{tab.1} for the detailed information of the two systems.

The power law relationships between $F$ and $P$ in two systems, which can be expressed as

\begin{equation}
\label{eq.1}
{F}\sim{P}^{\gamma}
\end{equation}
or
\begin{equation}
\label{eq.2}
Log(F)=\gamma Log(P)
\end{equation}

are found. Orthogonal regression \cite{14} is used to estimate the value of $\gamma$ in eq.~(\ref{eq.2}). We use orthogonal regression but not ordinary least-squares regression because the latter tends to overstate the effect of outliers (data points with large variance). It is observed that the empirical values of growth rate $\gamma$ in both systems are greater than 1. This finding of accelerating growth is consistent with results of previous studies \cite{3}\cite{4}\cite{5}\cite{6}. According to eq.~(\ref{eq.1}), it is easy to know that there is also a power law relationship between the average number of tags M posted by a user and the population, since $M=F/P\sim{P}^{\gamma-1}\sim{P}^{\theta}$. As $\gamma>1$ and $\theta>0$ in the data (Table~\ref{tab.1}), the average number of tags is not a constant; instead, it increases with population.

Comparing the $\theta$ in data with those reported in previous studies sheds lights on the accelerating nature of human collective behavior \cite{7}\cite{8}\cite{9}\cite{15}. The phenomenon that average activities (e.g., average salary or average walking speed of pedestrians) increase with city population has been well studied \cite{7}\cite{15}, and is used to explain the different paces of urban and rural life. Similar patterns also include GDP per capita that increases with country size \cite{8} and the average length of references in publications that increases with the scale of scientific collaboration network \cite{9}. However, most of the $\theta$ observed offline are not greater than their counterparts online. For instance, the average activity usually scales to the system size with a $\theta$ range from 0 to 0.35 in offline systems \cite{7}\cite{8} whereas $\theta$ found in my study can be greater than 0.38 (in Flickr). Given the population of a system, a larger $\theta$ means higher productivity in generating activities. Therefore we are reasonable to conjecture that online social systems, if are utilized appropriately, may be more ``productive" than offline ones.

Then, why is there accelerating growth and what determines the value of $\theta$ (or $\gamma$ )? Although several models have been proposed to answer the questions \cite{16}\cite{17}\cite{18}, there is still lack of a unified framework explaining the origin of accelerating growth. In the current study, I find that the accelerating growth relates to the stable (or time-invariant) heterogeneity in individual activities. In the next section, I will analytically show how heterogeneity gives rise to accelerating growth. Please note that this study does not aim to give an exclusive and unified framework towards the origins of all accelerating growth patterns, although I hope my model may contribute to such a framework in one way or another.

\section{Heterogeneity and accelerating growth}
\label{sec.3}

Many models have been proposed to explain the nonlinear growth phenomenon \cite{16}\cite{17}\cite{18}\cite{19}, but few can be generalized across different fields of study. Some ecologists suggest that nonlinear growth originates from the self-similar network structure of biological individuals \cite{17} or metabolic networks that minimize transportation cost \cite{18}. Scholars in linguistics suggest that a skewed distribution of word occurrence leads to non-linear growth in word vocabulary \cite{16}. Similar ideas are also seen in studies on the growth of family name \cite{19} and open source software \cite{5}. In the current research, I analytically attribute accelerating growth of online tags to the time-invariant heterogeneity of individual tagging activities in the system. The heterogeneity is quantified as the $1/\beta$, in which $\beta$ is the exponent of power law distribution of individual activities. The positive correlation between heterogeneity $1/\beta$ and accelerating growth rate $\gamma$ is verified by empirical data and numerical simulations.

\subsection{Time-invariant power law distribution in online tagging systems}
\label{sec.3.1}
\begin{center}
  \begin{figure*}[!ht]
  \centering
      \subfloat[\label{subfig-1:b}]{%
      \begin{overpic}[scale=0.5]{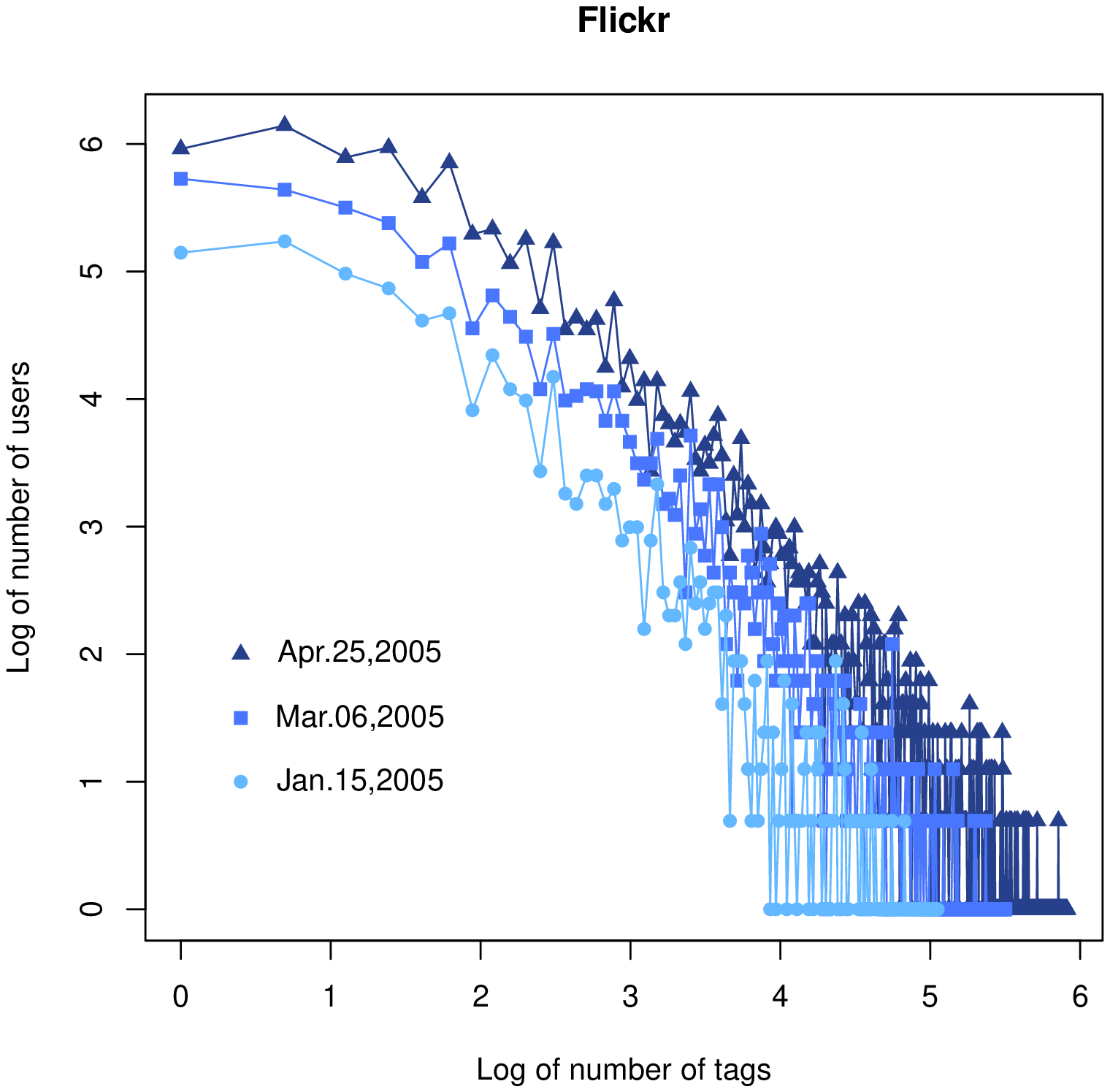}
        \put(61,56){{\includegraphics[scale=0.15]
{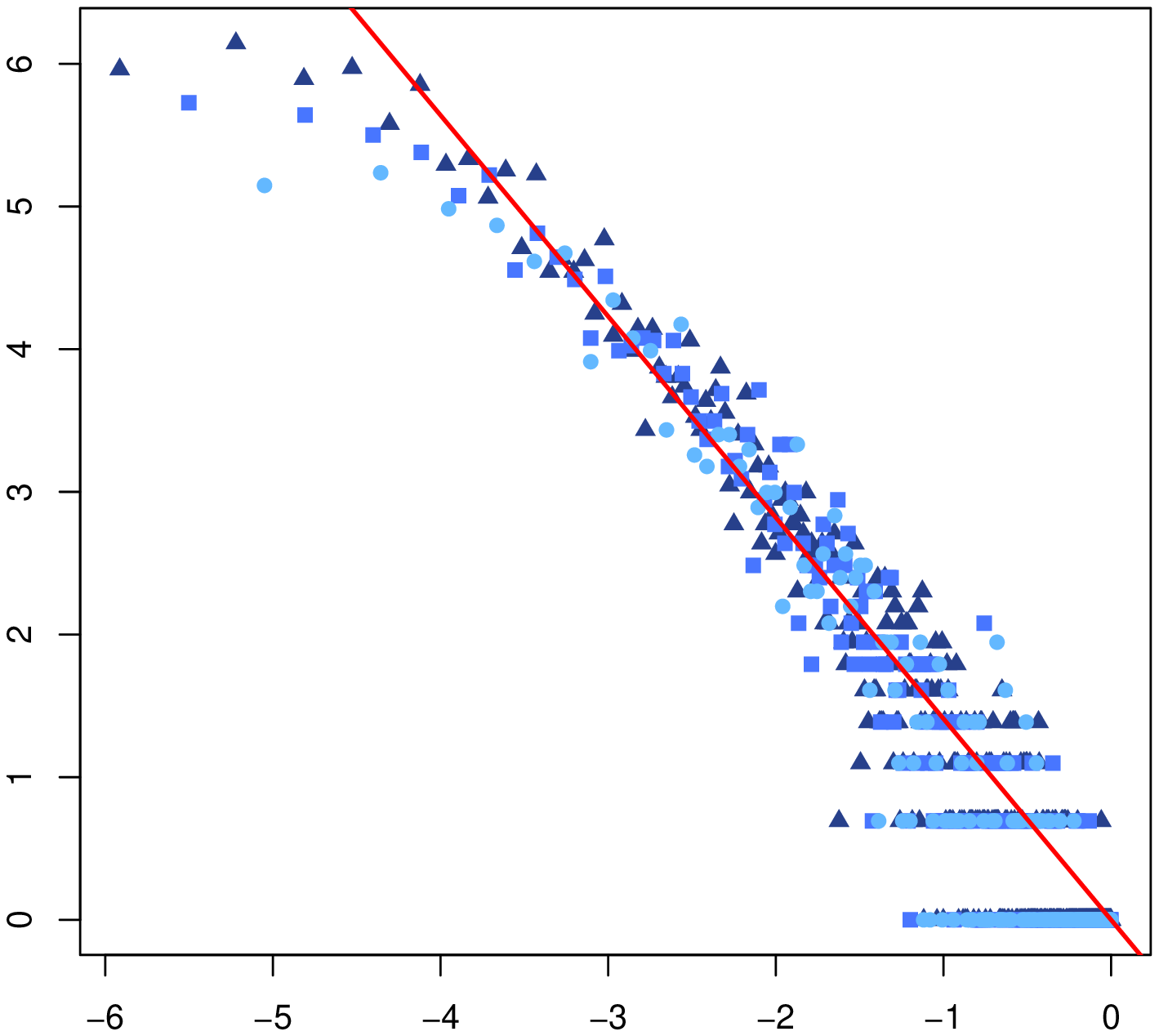}}}
      \end{overpic}
    }
       \subfloat[\label{subfig-2:a}]{%
      \begin{overpic}[scale=0.5]{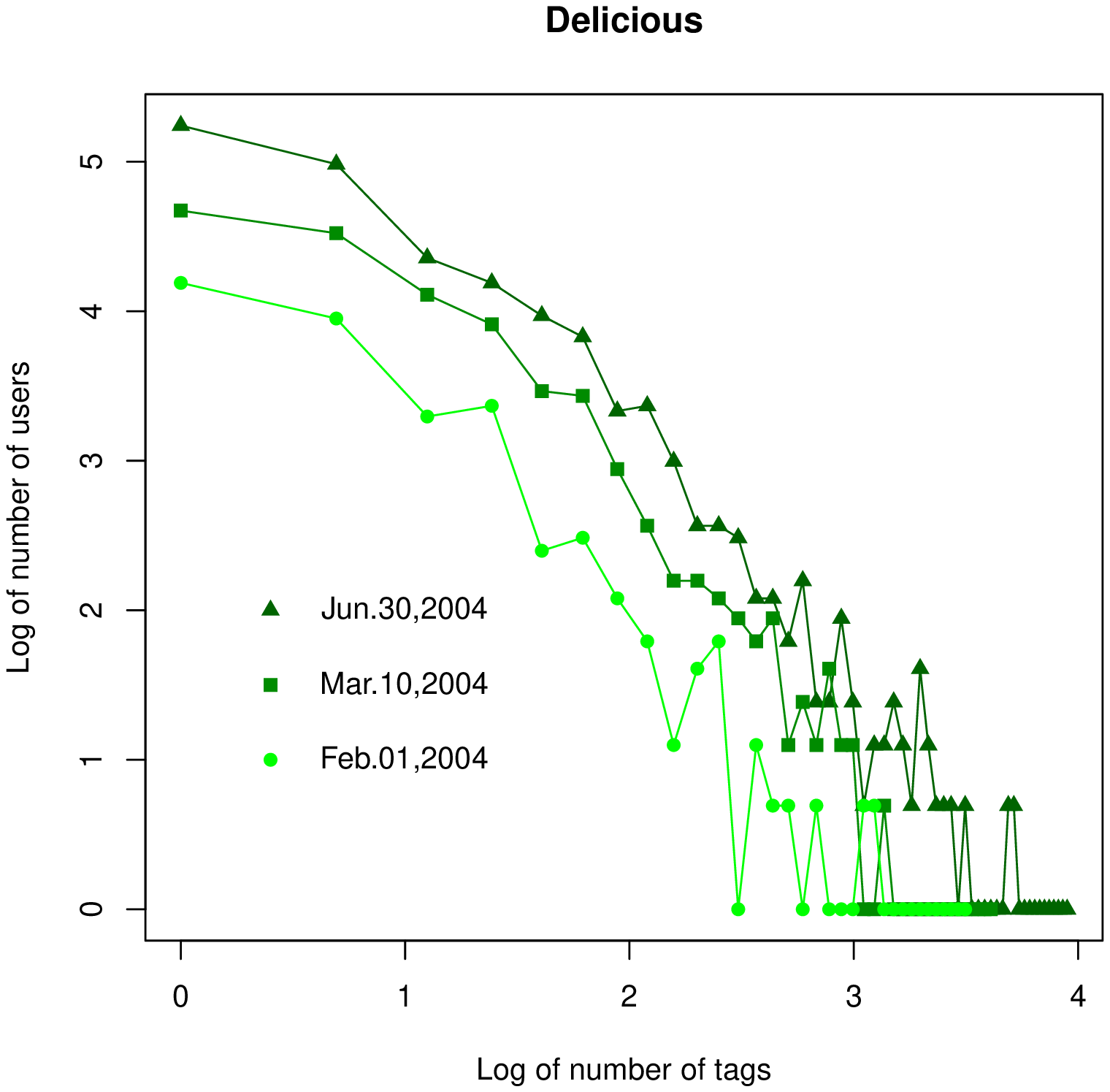}
        \put(61,56){{\includegraphics[scale=0.15]
{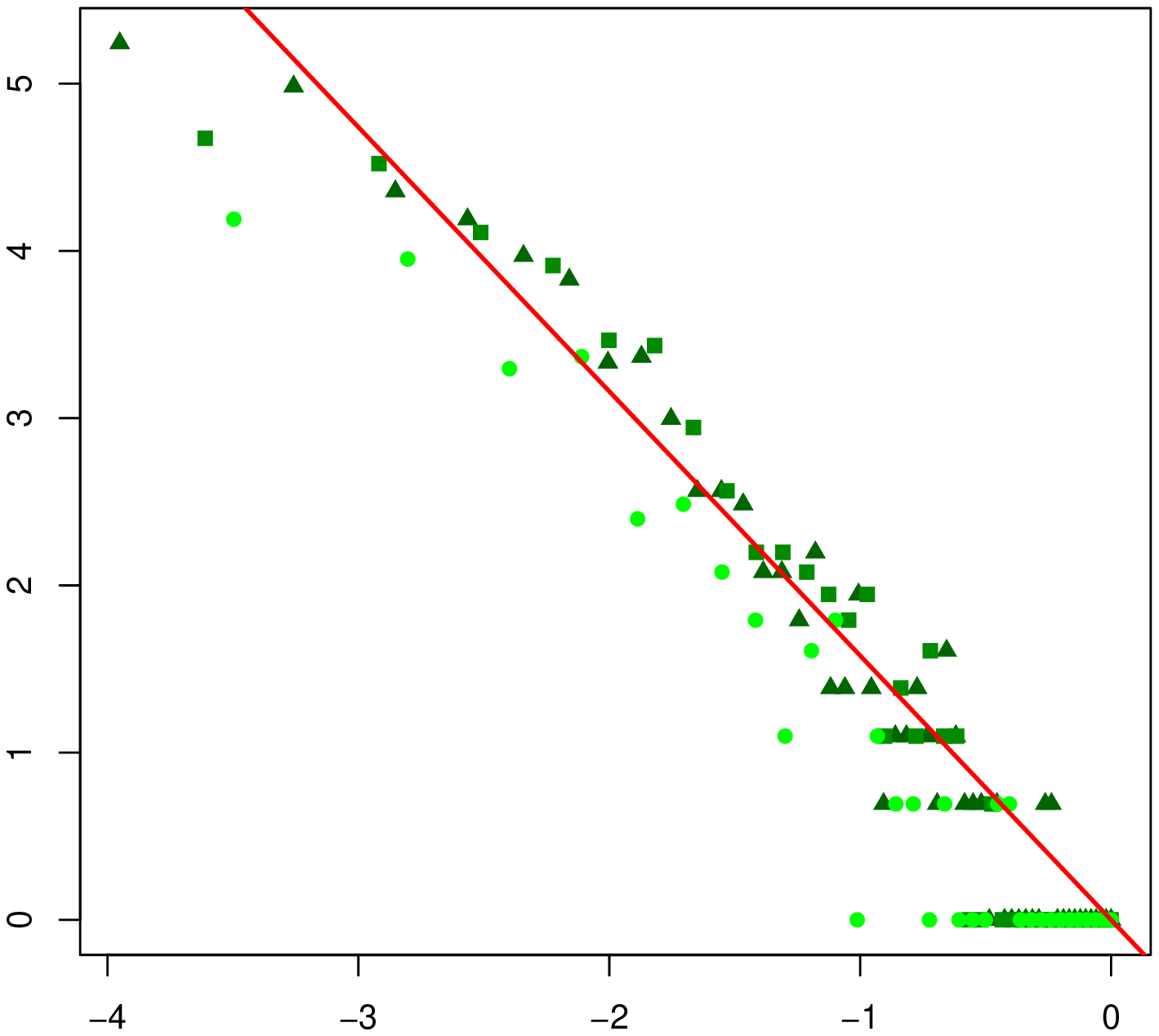}}}
      \end{overpic}
    }
\caption{Three examples of daily distributions of user tagging activities. The x axis is the log of number of tags and the y axis is the log of number of users. The rescaled distributions and the theoretical power law models with $\beta=1.41$ (Flickr) and $\beta=1.58$ (Delicious) are shown in the insets.}
    \label{fig.2}
  \end{figure*}
 \end{center}

To find out why the average number of tags increases with population, I explore the daily distributions of user tagging activities. It turns out the distributions approach a straight line in log-log axis, suggesting a model of power law distribution. I use $f$ to denote number of tags and $n(f)$ the number of users that generate so many tags. Then the power law probability distribution function (PDF) of $f$ be expressed as

\begin{equation}
\label{eq.3}
n(f) = C_{t}f^{-\beta}
\end{equation}

In which $C_{t}$ varies over time. In Eq.3 the value of $\beta$ is set to be greater than one, since the cumulative distribution function (CDF) of power law distribution (Eq.3), called Pareto distribution \cite{20}, is also a power law distribution with an exponent -$\alpha=1-\beta<0$.

In Figure 2, I draw 3 samples of daily distributions of individual activities on log-log axes for both systems. It is observed that, when population increases, the distributions move toward the right hand in parallel. Hence, I conjecture that $C_{t}$ in eq.~(\ref{eq.3}), which determines the intercept of the regression line, increase with system size, but $\beta$, which decides the slope of the line, is a size invariant constant. In other words, all the daily distributions should collapse to one theoretical curve if the variance of system size is controlled. I call the theoretical curve time-invariant power law distribution, since its slope does not change over time.

What makes the system maintain a time-invariant power law distribution in the temporal evolution? This is a question calling for further exploration. However, this phenomenon is non-trivial, since it means that the heterogeneity of the system does not change over time, which will give rise to accelerating growth. But before I discuss the connection between heterogeneity and accelerating growth, I would like to derive the function of time-invariant distribution at first.

As in tagging systems, there is usually only one user who generates the maximum activity, that is

\begin{equation}
\label{eq.4}
n(f_{max}) = 1 \approx C_{t}f_{max}^{-\beta}\rightarrow C_{t} = f_{max}^{\beta}
\end{equation}

With eq.~(\ref{eq.4}), we can rewrite eq.~(\ref{eq.3}) as

\begin{equation}
\label{eq.5}
n(f) = (f/f_{max})^{-\beta}
\end{equation}

eq.~(\ref{eq.5}) is the function of time-invariant power law distribution. As mentioned, all empirical daily distributions are supposed to collapse to the curve predicted by eq.~(\ref{eq.5}) after rescaling. Therefore in both systems under study I rescale daily distributions together before estimating the values of $\beta$ . According to the large values of adjusted $R^{2}$ shown in Table~\ref{tab.2}, the assumption about time-invariant distribution is plausible.

\begin{table}\footnotesize
\caption{Parameters in time-invariant power law model of online tagging systems.}
\label{tab.2}
\begin{center}
\begin{tabular}{lccccc}
\hline
  System & $\beta$ & 95$\%$ CI of $\beta$ & Adjusted $R^{2}$ & N of Days \\  \hline
  Delicious & 1.58 & [1.57, 1.59] & 0.90 & 171 \\
  Flickr  & 1.41 & [1.40, 1.41] & 0.90 & 120 \\ \hline
\end{tabular}
\end{center}
\end{table}

\subsection{From power law distribution to accelerating growth}
\label{sec.3.2}

\begin{figure}
\resizebox{1\columnwidth}{!}{
  \includegraphics{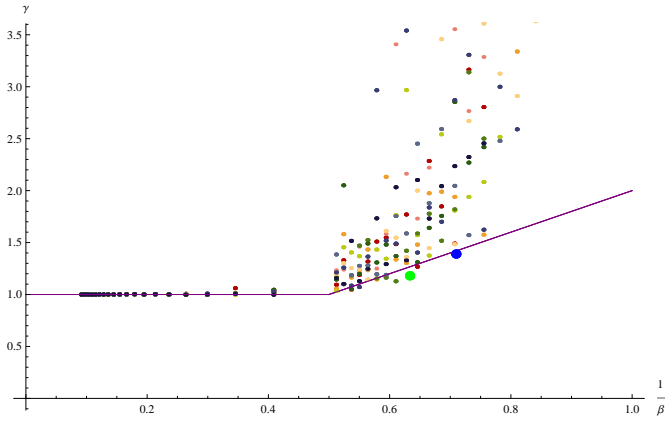}
}
\caption{The relationship between heterogeneity ($1/\beta$) and growth rate ($\gamma$). The analytical function is draw in purple line, the results of numerical simulations are plotted in small dots whose color varies with $C$, and the empirical results of two tagging systems are marked in green (Delicious) and blue (Flickr) dots with large size. In the numeric simulation based on power law PDF $f(x) = C^{\beta-1}x^{-\beta}(\beta-1))$, $C$ is an integer in $1\le C \le 10$, and $\beta$ is a fraction in ($1<\beta\le10$). For each value of $C$, 40 different values of $\beta$ are selected. Therefore, there are 400 small dots in the figure.}
\label{fig.3}
\end{figure}

Given a distribution function, the relationship between the population $P = \int n(f)df $ and the total number of tags  $F = \int n(f)fdf $ is fully predictable. With the time-invariant power law distribution shown above eq.~(\ref{eq.5}), the time-invariant relationship between $P$ and $F$ can be derived as follows.

As $F_{max}\gg1$ in empirical data,

When $\beta>2$,
\begin{equation}
\label{eq.6}
P=\int_{1}^{f_{max}} \!n(f)\,\mathrm{d}f\\=\frac{f_{max}}{1-\beta}-\frac{f_{max}^{\beta}}{1-\beta}\approx\frac{f_{max}^{\beta}}{\beta-1}
\end{equation}
\begin{equation}
\label{eq.7}
F=\int_{1}^{f_{max}} \!n(f)f\, \mathrm{d}f\\=\frac{f_{max}}{2-\beta}-\frac{f_{max}^{\beta}}{2-\beta}\approx\frac{f_{max}^{\beta}}{\beta-2}
\end{equation}
Therefore $F\sim P$, $\gamma\approx1$, $\theta\approx0$. The greater $\beta$ is, the more closely $\gamma$ approaches to 1.

When $\beta=2$,
\begin{equation}
\label{eq.8}
P=\int_{1}^{f_{max}} \!n(f)\ \approx f_{max}^{2} \approx \int_{1}^{f_{max}} \!n(f)f\ =F
\end{equation}
Therefore $F\sim P$, $\gamma\approx1$, $\theta\approx0$.

Similarly, when $1<\beta<2$, we can derive that
\begin{equation}
\label{eq.9}
P \approx \frac{f_{max}^{\beta}}{\beta-1}
\end{equation}
\begin{equation}
\label{eq.10}
F \approx \frac{f_{max}^{2}}{2-\beta}
\end{equation}
Therefore $F\sim P^{\frac{2}{\beta}}$, $\gamma\approx\frac{2}{\beta}$, $\theta\approx\frac{2}{\beta}-1$. And smaller $\beta$ leads to greater $\gamma$.

To sum up, we get
\begin{eqnarray}
\label{eq.11}
\gamma = \left\{ \begin{array}{rl}
 2/\beta &\mbox{ if $1<\beta<2$ } \\
  1 &\mbox{ if $\beta\ge2$}
       \end{array} \right.
\end{eqnarray}

According to eq.~(\ref{eq.11}), if a system maintains a time-invariant power law distribution with an exponent $\beta<2$, it will show property of accelerating growth. As power law distribution is famous for its high heterogeneity (i.e., quantitative difference in individual activities in the system), and smaller $\beta$ means higher heterogeneity, we can regard $1/\beta$ as an index for heterogeneity and claim that greater heterogeneity gives rise to a faster growth (greater $\gamma$). As shown in Fig.~\ref{fig.3} and Table~\ref{tab.3}, the theoretical function of $\gamma$ on $1/\beta$ (eq.~(\ref{eq.11})) is consistent with empirical data and simulation results.

\begin{table}\footnotesize
\caption{ Comparison between empirical ($\gamma$) and theoretical ($\gamma'$) accelerating growth rate}
\label{tab.3}
\begin{center}
\begin{tabular}{lccccc}
\hline
  System & $\beta$ & $\gamma'=2/\beta$ & 95$\%$ CI of $\gamma$ \\  \hline
  Delicious & 1.58 & 1.27 & [1.11, 1.31] \\
  Flickr  & 1.41 & 1.42 & [1.32, 1.46] \\ \hline
\end{tabular}
\end{center}
\end{table}

\section{Discussions}
\label{sec.4}

In this paper, I discuss the accelerating growth of two online tagging systems, noting that it is the time-invariant heterogeneity of individual tagging activities that gives rise to such a accelerating growth. Further, I analytically derive the function of accelerating growth rate $\gamma$ on heterogeneity $1/\beta$ , which is verified by empirical data and numerical simulations.

The findings of this paper support non-linear models of collective human behavior, which has been widely acknowledged by previous studies \cite{2}\cite{3}\cite{4}\cite{5}\cite{6}. Moreover, the relationship between heterogeneity and accelerating growth may not only exist in online tagging systems, but also can be found in other online social systems, or even offline social systems.

Accelerating growth, once proved to be widely exist in online social systems, can find its applications in website management or semantic web analysis. For instance, by analyzing web traffic data, we can predict the traffic on a website with a given active population. Hence this pattern will help website masters plan their web server capacity accordingly. Also, we can compare $\gamma$  between websites with equivalent functions and benchmark the most efficient one for other websites.

Some questions left points out the direction for the future studies. For example, why can online tagging systems maintain a time-interval power law distribution? Why $C_{t}$ in eq.~(\ref{eq.3}) is only determined by system size? These are all interesting questions worth further exploration.

\def\ack{\section*{Acknowledgements}%
    \addtocontents{toc}{\protect\vspace{6pt}}%
    \addcontentsline{toc}{section}{Acknowledgements}%
}
\ack{The author thanks Jonathan J. H. Zhu for providing comments on an earlier version of this paper.}

\end{document}